\documentclass[12pt]{article}
\pdfoutput=1
\usepackage{tikz}
\usetikzlibrary{decorations.pathmorphing}
\usepackage{putex}
\usepackage{feyn}
\usepackage[vcentermath]{youngtab}
\usepackage{subfig,braket}
\usepackage{lscape}
\usepackage{indentfirst}
\usepackage{graphicx}
\usepackage{epstopdf}
\usepackage{enumerate}
\usepackage{cite}
\usepackage{tensor}
\usepackage{slashed}
\usepackage{amsmath}
\usepackage{amssymb}
\usepackage{mathrsfs}
\usepackage{lgrind}

\usepackage{bbm}

\usepackage{hyperref}

\numberwithin{equation}{section}

\newcommand {\be} {\begin {equation}}
\newcommand {\ee} {\end {equation}}

\newcommand {\bes} {\begin {equation*}}
\newcommand {\ees} {\end {equation*}}

\newcommand{\eps}{\epsilon}

\def\CH{{\cal H}}

\newcommand{\beq}{\begin{equation}}
\newcommand{\eeq}{\end{equation}}

\def\be{ \begin{equation} }
\def\ee{ \end{equation} }

\def \be {\beta}

\def \beq { \begin{equation}}
\def \eeq {\end{equation}}

\renewcommand\Im{\operatorname{Im}}

\begin{document}


\institution{PU}{Department of Physics, Princeton University, Princeton, NJ 08544}
\institution{NYU}{CCPP, Department of Physics, NYU, New York, NY,  10003, USA}

\authors{Alexander M. Polyakov\worksat{\PU}, Fedor K. Popov\worksat{\NYU}}

\title{Kronecker anomalies and gravitational striction\footnote{Dedicated to Prof. Chen Ning Yang with admiration and awe.}}

\abstract{
We study quantum field theories in which the number of degrees of freedom changes discontinuously across the momentum space. This discontinuity which we call "Kronecker anomaly" leads to non-local effective actions and can be represented as a theory with the random, self-tuning coupling constants.
}

\maketitle

\tableofcontents

\section{Introduction and General Discussion}
One of the most surprising features of our world is that it is described by local quantum field theories (QFT). This means that we associate with the object we want to describe (be it the standard model or a classical magnet) a set of local operators $\left\{O_n(x)\right\}$ satisfying operator product expansion \cite{wilson1969non,kadanoff1969operator,polyakov1969properties}
\begin{gather}
        O_n \cdot O_m = \sum c^l_{nm} O_l. \label{eq:OPE}
\end{gather}
This also means, that small perturbation can be described by the conformal perturbation theory, $V=\sum_a \lambda_a O_a$, where $\left\{\lambda^a\right\}$ are the coupling constants. In the general QFT, the operator algebra \eqref{eq:OPE} replaces the equations of motion. But is the locality a must? We don't know the answer. 

In this paper we show that there is a plausible mechanism discovered by Larkin and Pikin \cite{larkin1969zh,pikin1993weak}, which breaks the locality. It was already mentioned long ago \cite{polyakov1991self}, but many related facts remain to be clarified. 

The effect which we discuss and generalize in this paper is the following. Consider a $\phi^4$ theory on a crystal lattice near a critical point. The effective action is given by
\begin{gather}
    \mathcal{L}_ 0 = \frac12 \left(\partial_\mu \phi\right)^2 + \tau \phi^2 + g \phi^4, \label{eq:scalphi4}
\end{gather}
where $\tau \propto \left(T-T_{\rm cr}\right)$ and $\phi$ is magnetization. As we account for the lattice vibrations we get another field, the displacement $u_\alpha(x)$ and strain $u_{\alpha\beta} = \partial_\alpha u_\beta + \partial_\beta u_\alpha$.  The Lagrangian for this field is given by
\begin{gather}
    \mathcal{L}_1 = \frac{\lambda}{2} u_{\alpha\alpha}^2 + \mu u_{\alpha\beta}^2, \label{eq:disaction}
\end{gather}
where $\lambda,\mu$ are Lame's coefficients. Finally there is an interaction leading to magneto-striction
\begin{gather}
    \mathcal{L}_2 \propto u_{\alpha\alpha} \phi^2. \label{eq:perturbation}
\end{gather}
The field $u_{\alpha\beta}$ is Gaussian and if we naively integrate out $u_{\alpha\beta}$, the result seems trivial. Indeed, if we use \eqref{eq:disaction} we get
$$
\braket{u_{\alpha\alpha}(x) u_{\beta\beta}(y)} \propto \delta^d(x-y)
$$
and the dressing of $\phi$ with $u$ fluctuations results with
$$
\Delta \mathcal{L}_{\rm eff} \propto \int \braket{u_{\alpha\alpha}(x) u_{\beta\beta}(y)} \phi^2(x) \phi^2(y) d^d x d^d y  \propto \int \phi^4(x) d^d x
$$
So, we obtain a trivial renormalization of the $\phi^4$ term.

There is a flaw in this calculation since $u_{\alpha\beta}(x)$ are not independent fields but are subject to the Saint-Venant compatibility conditions \cite{landau1986theory}
\begin{gather}\label{RicciLin}
R_{\alpha\beta}=\partial^2 u_{\alpha\beta} +\partial_\alpha \partial_\beta u_{\gamma\gamma} - \partial_\alpha \partial_\gamma u_{\gamma\beta} - \partial_\beta \partial_\gamma u_{\gamma\alpha}=0
\end{gather}
which ensures that for $k \neq 0$ we can express $u_{\alpha\beta}$ in terms of some field $u_\alpha$
$$
u_{\alpha\beta} = k_\alpha u_\beta + k_\beta u_\alpha
$$
The key point of the present paper is that the statistics of $u_{\alpha\beta}(k)$ changes discontinuously at $k=0$, because at this point  Saint-Venant's condition disappears. This jump of the numbers of the degrees of freedom leads to a non-local contribution of the elastic modes proportional to the Kronecker symbol, $\delta_{\vec{p},0}$, which we call the Kronecker anomaly. To be more general and precise, consider a field theory with periodic boundary conditions or, which is the same, a theory on a $D$-dimensional torus. The Green functions in such a theory have the following structure
\begin{gather}
    G\left(\vec{p}_1,\ldots, \vec{p}_N\right)=
    \tilde{G}\left(\vec{p}_1,\ldots, \vec{p}_N\right) + A(0,\vec{p}_2,\ldots, \vec{p}_N) \delta_{\vec{p}_1,0} + \ldots, \label{eq:KAdef}
\end{gather}
where $\tilde{G}\left(\vec{p}_1,\ldots,\vec{p}_N\right)$ is assumed to be an analytic function of its arguments\footnote{We also assume that these functions do not have essential singularities.}.  Then s non-zero contribution $A(0,\vec{p}_2,\ldots,\vec{p}_N)$ is a Kronecker anomaly. We can formulate it differently --- the Green functions on a torus could not be analytic functions due to the Kronecker anomaly. 

To proceed, let us find the propagator for the acoustic field $u_{\alpha\alpha}$ which enters formula \eqref{eq:perturbation}. For $\vec{k}\neq 0$ we use \eqref{eq:disaction} and the representation $u_{\alpha\beta} = k_\alpha u_\beta + k_\beta u_\alpha$. This part of the action has the form
\begin{gather}
S = \sum_{\vec{k} \neq 0} \left( \mu k^2 \delta_{\alpha\beta} + \left(\lambda + \mu\right) k_\alpha k_\beta\right) u_\alpha(k) u_\beta(-k) = \notag\\
= \sum_{\vec{k}\neq 0} \left( \mu k^2 \left(\delta_{\alpha\beta } - \frac{k_\alpha k_\beta}{k^2} \right) + \left(\lambda + 2 \mu\right) k^2  \frac{k_\alpha k_\beta}{k^2} \right). \label{eq:kneq0}
\end{gather}

These two pieces are orthogonal and the propagator, being the inverse of the above matrices is given by
\begin{gather}
\braket{u_\alpha(k)u_\beta(-k)}_{k \neq 0} = \frac{1}{\mu k^2}\left[ \left(\delta_{\alpha\beta} - \frac{k_\alpha k_\beta}{k^2} \right) + \frac{1}{\lambda + 2\mu} \frac{k_\alpha k_\beta}{k^4}\right]. \label{eq:prop}
\end{gather}
So, the $k \neq 0$ part of $u_{\alpha\alpha}$ is simply
\begin{gather}
\braket{u_{\alpha\alpha}(k) u_{\beta\beta}(-k)} = \frac{1}{\lambda + 2 \mu} \quad \text{at $k \neq 0$} \label{eq:propu},
\end{gather}
For $\vec{k} =0$ the propagator is given by \footnote{We note that these modes $\vec{k}=0$ could not be present in the spectrum depending on the structure of the crystal \cite{golubovic2000structural}.}
\begin{gather}
\braket{u_{\alpha\beta} u_{\gamma\delta}} \propto \int \mathcal{D} u e^{-\beta E(u)} u_{\alpha\beta} u_{\gamma\delta} ,
\end{gather}
where $E(u)$ is again the elastic energy \eqref{eq:disaction}. Using Ward's identity
\begin{gather}
\left< \frac{\delta E(u)}{\delta u_{\alpha \beta}} u_{\gamma\delta} \right> = \delta_{\alpha \gamma} \delta_{\beta\delta} + \delta_{\alpha\delta}\delta_{\beta\gamma}, \notag\\
\frac{\delta E(u)}{\delta u_{\alpha \beta}} = \left(\lambda + \frac{2\mu}{d} \right) u_{\gamma\gamma} \delta_{\alpha\beta} + 2 \mu \left(u_{\alpha \beta} - \frac{1}{d} \delta_{\alpha\beta} u_{\gamma\gamma} \right), \braket{u_{\alpha\alpha} u_{\beta\beta}} = \frac{1}{\lambda + \frac{2\mu}{d} }
\end{gather}
Now we have the propagator
\begin{gather}
\braket{u_{\alpha\alpha}(k) u_{\beta\beta}(-k)} = C_1 + C_2 \delta_{\vec{k},0}
\end{gather}
to which in the coordinate space has the form
\begin{gather}
\braket{u_{\alpha\alpha}(x) u_{\beta\beta}(y) } = C_1 \delta(x-y) + \frac{C_2}{V},
\end{gather}
where $V$ is the volume and
\begin{gather}
C_1 = \frac{1}{\lambda+ 2\mu} , \quad C_2 = \frac{2 \left(1 - \frac{1}{d}\right) \mu}{\left(\lambda + 2 \mu\right)\left(\lambda + \frac{2}{d}\mu\right)}   \label{eq:c1c2}
\end{gather}
the term $C_2$ represents the Kronecker's anomaly \eqref{eq:KAdef}. Since the $u$-field is Gaussian, it is easy to calculate its effect on the partition function of the order parameter
\begin{gather}
\int \mathcal{D} u e^{-S[u_{\alpha\beta}]}\exp\left[\int d^d x u_{\alpha\alpha}(x) \phi^2(x)\right]   = \exp\left[\int dx dy \braket{u_{\alpha\alpha}(x) u_{\beta\beta}(y)} \phi^2(x) \phi^2(y) \right]. \label{eq:gausinter}
\end{gather}
Thus, the partition function with the account of elasticity acquires an extra term coming from \eqref{eq:gausinter}
\begin{gather}
Z = \int \mathcal{D}\phi \exp\left\{\int d^d x\left[\left(\nabla \phi\right)^2 + \tau \phi^2 + \bar{g} \phi^4 \right] + \frac{C_2}{V} \int \phi^2(x) \phi^2(y) d^dx d^dy \right\} = \notag\\
=\int \mathcal{D}\phi \mathcal{D}\xi \exp\left\{\xi^2 V + C_2^{-1} \int d^d x (\tau + \xi) \phi^2 \right\}.
\end{gather}
So, we see that the coupling of the order parameter to acoustic phonons leads to the bilocal term in the action
\begin{gather}
\Delta S \sim \frac{1}{V} \int \eps(x_1) \epsilon(x_2) d^d x_1 d^d x_2,
\end{gather}
the factor $\frac{1}{V}$ is needed to make this additional term relevant in the thermodynamic limit. This term can also be cast in the form
\begin{gather}
\Delta S \sim \xi^2 V + \int d^d x \xi \eps(x),
\end{gather}
where $\epsilon(x) \sim \phi^2(x)$ is the energy density of the magnet. The $\xi$ variable plays the role of the coupling constant and we see that the coupling constants behave as random Gaussian variables. 

Quite unexpectedly, the random coupling constants and bilocal operators have been discussed long ago in a total different context. In the works \cite{coleman1988there,hawking1990wormholes} it was argued that the wormholes and baby universes generate bilocal perturbations.
In our set up, one can obtain random coupling much cheaper, we need just an acoustic phonon instead of the other universe.

But this it not the end of the story. Depending on the values of the anomalous dimensions, the integral (17) may  diverge. Using the OPE
\begin{gather}
    \epsilon(x_1) \epsilon(x_2) = \frac{1}{|x_1 - x_2|^{2\Delta_\epsilon}} I + \ldots \label{eq:opeeps},
\end{gather}
we find that if $\Delta_\epsilon< \frac{d}{2}$ we have the bilocal action and coupling are random, while for $\Delta_\epsilon>\frac{d}{2}$ we have UV-divergence and the bilocal action disappears.

The above discussion shows that the essence of bilocality lies in Kronecker's terms. An important example of this mechanism is the cosmological constant problem. In this case the role of the acoustic phonons is played by gravitons. We have a metric tensor, $g_{\mu\nu} = \eta_{\mu\nu} + h_{\mu\nu}$, with the gauge condition
\begin{gather}
k_\mu h_{\mu \nu} = \frac12 k_\nu h_{\lambda\lambda}
\end{gather}
and it is obvious that we have extra degrees of freedom at $k=0$, where the gauge condition degenerates. However, a more subtle analysis of gauge invariance is needed for the cosmological constant problem. We leave this problem for the future work and present here only a simpler case of the 2d string theory in the dilaton background \cite{polchinski1998string,Klebanov:1991hx,Polyakov:1991xa}. In this case we have two sets of oscillators satisfying the relation
$$\left[a^\mu_n ,a^\nu_m\right] = n \eta^{\mu\nu} \delta_{n,-m},$$
and all states could be build by applying these operators at vacuum state $\ket{0}$. 
Naively, there are no photon-like degrees of freedom in two dimensions. However, due to the Kronecker anomaly, infinite number of the so called discrete states appear in the theory. As discussed in \cite{Klebanov:1991hx,Polyakov:1991xa,witten1992ground,polchinski1998string}, the first excited state of the open string has the usual form $e_\mu(p) a^\dagger_{1\mu} \ket{0}$ and satisfy the gauge condition
\begin{gather}
p(p+2b) = 0, \quad \left(p_\mu + 2 b_\mu\right) e_\mu(p) = 0, \quad e_\mu \sim e_\mu + \lambda p_\mu,
\end{gather} 
where $b_\mu$ is proportional to the  derivative of the dilaton $b_\mu \sim \partial_\mu \varphi$, $\varphi$ is  the dilaton. This quantity is fixed by the condition that the central charge of the string is zero. In $1+1$ dimension we can try to eliminate the "photon" $e_\mu(p)$. We choose a gauge and set $e_1=0$ and then using the constraint to get $e_0 = 0$. This is a textbook trick to show that in $d$-dimensions photon has $(d-2)$ degrees of freedom. In the present case we also can do this except of the points $p_\mu = 0$ and $p_\mu = -2 b_\mu$. Analogous reasoning works for the higher string excitations. As a result our string has infinite number of the so called discrete states. This was explicitly verified in \cite{Klebanov:1991hx,Polyakov:1991xa} by calculating scattering amplitudes, which contain infinite number of poles. Their very non-trivial dynamics is important for establishing string-matrix model correspondence.

We will not discuss here the dynamics of discrete states, but just present the second level string excitation, which was discussed in \cite{Klebanov:1991hx,Polyakov:1991xa,gross1991two}. It has the form
\begin{gather}
\ket{\psi_2} = \left(e_\mu a^\dagger_{2\mu} + \xi_{\mu\nu} a^\dagger_{1\mu} a^\dagger_{1\nu} \right)\ket{p}.
\end{gather}
The coefficients $e_\mu, \xi_{\mu\nu}$ satisfy the constraint
\begin{gather}
\left(p_\mu + 2 b_\mu\right) \xi_{\mu\nu} + e_\nu =0 ,\quad \left(p_\mu + 3 b_\mu\right) e_\mu + \frac12 \xi_{\lambda\lambda} = 0,
\end{gather}
and two gauge transformations
\begin{gather}
e_\mu \sim e_\mu + \eta_\mu, \quad \xi_{\mu\nu} \sim \xi_{\mu\nu} + \frac{1}{2}\left(p_\mu \eta_\nu + p_\nu \eta_\mu\right),\quad \left(p_\mu + 2 b_\mu\right) \eta_\mu=0, \notag\\
e_\mu \sim e_\mu + \lambda\left(5 p_\mu -2 b_\mu\right),\quad \xi_{\mu\nu} \sim \xi_{\mu\nu} + \lambda\left(\delta_{\mu\nu} + 3 p_\mu p_\nu\right) 
\end{gather}
The analysis in \cite{Klebanov:1991hx,Polyakov:1991xa} shows that the physical states exist only at $p_1 =\pm \frac12 $, as a result of the degeneration of the gauge condition. Deeper dynamical understanding of this system seems highly desirable.

\section{Kronecker Anomaly in Thermal Harmonic Oscillator}
The other way to generate non-local terms in the effective action is to consider a theory at thermal state (or on a torus). For instance, we can study a simple harmonic oscillator at temperature $\beta$ with the action
\begin{gather}\label{SHOaction}
S = \int\limits^\beta_0 d\tau \left[\frac12 \dot{x}^2(\tau) +\frac{1}{2} m^2 x^2(\tau)\right],
\end{gather}
where we assume periodic boundary conditions $x(\tau+\beta)=x(\tau)$ and mass, $m$, plays the role of the frequency of the oscillator. From this we can derive the propagator in the Euclidian signature to be
\begin{gather}
G(\tau) = \frac{1+n}{2m} e^{-m |\tau|} + \frac{n}{2m} e^{m |\tau|},\,\, n = \frac{1}{e^{\beta m} - 1}, \,\, G(\omega_m) = \frac{1}{\omega_m^2 + m^2},\,\, \omega_m = \frac{2\pi m}{\beta},
\end{gather}
where $n$  was fixed by periodic boundary conditions. The action \eqref{SHOaction} is very simple and we can compute all physical observables, for instance, the susceptibility $\alpha=-\frac{\partial \braket{x^2}}{\partial m^2}$. It could be done by using Kubo formula $\alpha_\omega \propto \int\limits^\beta_0 d\tau e^{i\omega \tau} \braket{x^2(0) x^2(\tau)}$. First we do this computation in the lower-half plane of the frequency range $i\Omega \in \frac{2\pi}{\beta} \mathbb{N}$
\begin{gather}
\Pi(\Omega) = \begin{tikzpicture}[baseline={([yshift=-.5ex]current bounding box.center)},scale=2]
\draw[fill] (0,-1) circle [radius=1pt];
\draw[fill] (1,-1) circle [radius=1pt];
\node[left] at (0,-1) {$x^2$};
\node[right] at (1,-1) {$x^2$};
\draw (0,-1)..controls (0.5,0.25-1)..(1,-1);
\draw (0,-1)..controls (0.5,-0.25-1)..(1,-1);
\end{tikzpicture} =
T \sum_n \frac{1}{\omega_n^2 + m^2} \frac{1}{(\Omega-\omega_n)^2+ m^2} = \frac{1}{m} \frac{\coth \frac{\beta m}{2}}{4m^2 + \Omega^2}, \label{eq:KAupper}
\end{gather}
if one naively assumes that $\Pi(\Omega)$ is an analytic function of the frequency $\Omega$ we would get a wrong answer for the $\Pi(\Omega=0)$. To see this we make the same computation in the coordinate representation
\begin{gather}\label{directcomp}
\Pi(\tau) = G^2(\tau) = \frac{(1+n)^2}{4m^2} e^{-2 m |\tau|}  + \frac{n (1+n)}{ 2 m^2} + \frac{n^2}{4m^2} e^{2 m |\tau|},\notag\\
\quad \text{then}\quad \Pi(\Omega) =   \frac{1}{m} \frac{\coth \frac{\beta m}{2}}{4m^2 + \Omega^2}  + \frac{1}{4 m^2}\frac{1}{\cosh m \beta -1} \delta_{\Omega,0}, \label{eq:KAcomp}
\end{gather}
where the non-analytic term appears because of the interference between two exponents in the thermal propagator $G(\tau)$. It means that our calculation of the polarization operator in the lower-half plane of the frequency range does not capture the zero mode contribution in \eqref{eq:KAupper} as compared to \eqref{eq:KAcomp}. It actually plays an important role, because $\Pi(0)$ is a susceptibility and could be measured at the experiment. To sum up, we have two possible values for the susceptibility $\alpha$, that can be found by analytical continuation from the lower-half plane $\Pi_A$ and by direct computation of $\Pi_M$ at $\Omega=0:$
\begin{gather} \label{paradox}
\Pi_{A}(0) =  \frac{\coth \frac{\beta m}{2}}{4m^3},\quad \text{and}\quad
\Pi_{M}(0)= \frac{1}{m} \frac{\coth \frac{\beta m}{2}}{4m^2}  + \frac{1}{4 m^2}\frac{1}{\cosh m \beta -1}.
\end{gather}
To understand this discrepancy we compute the susceptibility for harmonic oscillator using the straightforward definition of $\alpha$ and the action \eqref{SHOaction},
\begin{gather*}
\braket{n|x^2|n} = \frac{2n+1}{2m}, \quad
\braket{\braket{x^2}} = \frac{\tr\left[ x^2\, e^{-\beta H}\right]}{\tr\left[e^{-\beta H}\right]} = \frac{1}{2\omega}\frac{\sum^\infty_{n=0}(2n+1) e^{-n \beta m}}{\sum^\infty_{n=0} e^{-\beta m}} = \frac{1}{2m}\coth \frac{\beta m}{2}.
\end{gather*}
We can change the mass ,$m$, in two different regimes: keeping temperature or entropy constant. They lead to two different types of the susceptibilities. First, we compute the isothermal susceptibility of the system,
\begin{gather}
\alpha_\beta=-\left(\frac{\partial \braket{\braket{x^2}} }{\partial m^2} \right)_{\beta}=  \frac{1}{m} \frac{\coth \frac{\beta m}{2}}{4m^2}  + \frac{\beta}{2 m^2}\frac{1}{\cosh m \beta -1} = \Pi_M(0),
\end{gather}
while, adiabatic or isentropic susceptibility is
\begin{gather}
\alpha_S = - \left(\frac{\partial \braket{\braket{x^2}} }{\partial m^2} \right)_{S}=   \frac{1}{m} \frac{\coth \frac{\beta m}{2}}{4m^2} = \Pi_A(0),
\end{gather}
Then from this computation we can understand the discrepancy in the quantities \eqref{paradox}. The computation of $\Pi_M(0)$ corresponds to the isothermal susceptibility because in the Matsubara technique the temperature is assumed to be fixed by the boundary conditions $x(\tau+\beta)=x(\tau)$. But we can show in general that $\Pi_A(\Omega)$ gives the answer for real-time computation \cite{landau1980statisticheskaia} with the use of the Keldysh-Schwinger technique
\begin{gather}
\Pi_{KS}(0) = \int\limits^t_{-\infty} dt' \operatorname{Im}\left[G_\beta^2(t-t')\right] dt' =  \frac{1}{m} \frac{\coth \frac{\beta m}{2}}{4m^2} =\Pi_A(0).
\end{gather}
It happens because for the real time with unitary evolution the entropy $S = - \tr\left[ \rho \log \rho \right]$ remains constant. 
Therefore, the discrepancy \eqref{paradox}  measures, roughly speaking, the difference between these two different susceptibilities, at constant temperature or entropy. 
According to our definition this discrepancy is the Kronecker anomaly \eqref{eq:KAdef}. The existence of the Kronecker term changes the behavior of the system. If we couple the oscillator to an external system and integrate its degrees of freedom, the resulted effective action would contain non-local terms for the external system as in the case of acoustic phonons. The same discrepancy could arise in quantum field theories.
For example, we can consider the following action
\begin{gather}
S =\int\limits^\beta_0 d\tau d x \left[\frac12 (\partial_\mu a)^2 + \frac12 (\partial_\mu \phi)^2 - \frac12 m^2 \phi^2 -g a \phi^2 \right].
\end{gather}
Then we can study one-loop correction to the propagator of the field $a$. It has the following form
\begin{gather}
G^{-1}\left(\Omega_n,p\right) = \Omega_n^2 + p^2 + m^2 - \Sigma(\Omega_n,p), \notag\\
\Sigma\left(\Omega,p\right) = g^2 T \sum_n \int \frac{d q}{2\pi}\frac{1}{\omega_n^2 + m^2 + \vec{q}^2}\frac{1}{(\Omega -\omega_n)^2 + m^2 + (p-q)^2}.
\end{gather}
We expect that $\Sigma(\Omega,p)$ has a non-analytical contribution at $p=0$ (then the particles will have the same energies and can contribute to the Kronecker anomaly \eqref{A:Step}).
And if we evaluate this function at $\Omega_n = 2\pi T n$  and continue to the $\Omega=0$ and compare it with the direct computation we would get two different answers:
\begin{gather}
\delta\Sigma = \Sigma(\Omega=0,\vec{p}=0)-\Sigma_A(0,\vec{p}=0) = \int \frac{d q}{2\pi}\frac{1}{8 T (m^2 + q^2)} \frac{1}{\sinh^2 \left[\frac{\sqrt{m^2 + q^2}}{2 T}\right]}
\end{gather}
This difference would induce an additional non-local term in the effective action. 

There are a lot of other interesting mechanisms that could lead to the emergence of bilocal actions. We mostly discuss them in the supplementary material. In appendix \ref{sec:KAE}, we discuss how the Kronecker anomaly could arise in the $U(1)$ gauge theory to the jump in the number of degrees of freedom for $k=0$ 
and $k\neq 0$.
In section \ref{sec:MAKA}, we show the physical meaning of this Kronecker anomaly that arise in a space with compact direction and give precise conditions for the emergence of it.  In appendix \ref{sec:KAELGR} we show how Kronecker anomaly could arise in quantum electrodynamics or gravity on a torus, coupled to a scalar or fermionic field.  In appendix \ref{sec:KA3}, we that in odd dimensional de Sitter spaces quantum field theories also have Kronecker anomalies, while in even dimensional de Sitters such non-analytical contributions are absent.  In appendix \ref{sec:Lacuna}, we show how discontinuous propagator in the coordinate space arise in the hyperbolic spaces.


\section*{Acknowledgment}

We are grateful to Boris Altshuler, Igor R. Klebanov, Juan M. Maldacena, Emil T. Akhmedov and P. Wiegmann for insightful discussions and suggestions throughout the project. F.K.P. is currently a Simons Junior Fellow at NYU and supported by a grant
855325FP from the Simons Foundation. 


\appendix

\section{Kronecker anomaly in electromagnetic theory} \label{sec:KAE}
The Kronecker anomaly appears also in the electromagnetic theory
\begin{gather}
    S = \frac{1}{4}\int d^d x F_{\mu\nu}^2,\quad F_{\mu\nu} = \partial_\mu A_\nu-\partial_\nu A_\mu,
\end{gather}
for $k \neq 0$ we can use Landau gauge $k^\mu A_\mu = 0$ to get the propagator
\begin{gather}
\braket{A_\mu(k) A_\nu(-k)} = \frac{g_{\mu\nu}-\frac{k_\mu k_\nu}{k^2}}{k^2}.
\end{gather}
The electromagnetic field is subjected to the following Maxwell equations
\begin{gather}
    \delta F = j,\quad dF=0,\quad \delta=*d*
\end{gather}
The first equation is dynamical, that shows how $F_{\mu\nu}$ depends on the external current, while the second equation is just a statement, that $F_{\mu\nu}$ is a closed form and for some $A_\mu, F = dA$, that is analogous to the equation \eqref{RicciLin}. Namely, if we consider the constraint with some wave vector $k^\mu$ this equation states that $F_{\mu\nu}(k)$ has $d-1$ degrees of freedom ($d$ comes from the consideration that $F=dA$ for some 1-form $A$ and $(-1)$ comes from gauge invariance of such a configuration $A\sim A + d\phi$ would give the same field strength $F$). 

At $k^\mu=0$, when constant configurations are considered we have $\frac{d(d-1)}{2}$ degrees of freedom (in 4 dimensions it corresponds to 6 constant electric and magnetic fields) and we would have the same type of the change of the number of degrees of freedom. Indeed, let us consider the following correlation
\begin{gather}
    \frac12\braket{F_{\mu\nu}(k) F_{\mu\nu}(-k)} = \frac{1}{k^4} \left(g_{\mu \nu} k^2 - k_\mu k_\nu\right)\left(g^{\mu \nu} k^2 - k^\mu k^\nu\right) = d-1,
\end{gather}
At $k^\mu=0$ the propagator is singular and naively it would give the same answer, $d-1$, but if we consider it more carefully we would get
\begin{gather}
    S = \frac{1}{4} F_{\mu\nu}^2 V,\quad  \frac12\braket{F_{\mu\nu}(k=0) F_{\mu\nu}(k=0)} = \frac{d(d-1)}{2} \frac{1}{V},
\end{gather}
giving us at the end
\begin{gather}
     \frac12\braket{F_{\mu\nu}(x) F_{\mu\nu}(y)} = (d-1) \delta^{d}(x-y) + \frac{(d-1)(d-2)}{2} \frac{1}{V}, 
\end{gather}
that is an analog of the equation \eqref{eq:c1c2}.
\section{Mathematical aspects of Kronecker anomaly} \label{sec:MAKA}
In this section we explain how the Kronecker anomaly arises in a loop computation. Let us consider a quantum field theory at finite temperature $\beta$ and try to compute any loop diagram.  For brevity we will omit the variable $p$ in the latter formulas. According to the Matsubara technique  we should sum over the discrete frequencies
\begin{gather}\label{ThermalSum}
\mathcal{A}_l(\omega_i) = T \sum_n G_M(i\Omega_n) G_M(i(\Omega_n+\omega_1))\ldots G_M(i(\Omega_n + \omega_1+ \ldots \omega_{l-1})),
\end{gather}
Since we have a quantum system at thermal equilibrium
\begin{gather}
G_M(i\omega_n) = \int \frac{d\omega}{2\pi} \frac{\Im G_R(\omega)}{\omega-i\omega_n},
\end{gather}
Generally, $\Im G_R(\omega)$ depends only on the spectrum of the model but not on the concrete state itself. For instance, for a harmonic oscillator $\Im G_R(\omega) = \pi \delta(\omega^2-m^2)$.   
We plug this representation in the amplitude \eqref{ThermalSum}
\begin{gather*}
\mathcal{A}_l(\omega_i) =  T \int \prod^l_{i=1}\frac{dx_i}{2\pi} \Im G_R(x_i) \sum_n \frac{1}{x_1 - i \Omega_n}\frac{1}{x_2 - i \Omega_n - i \omega_1} \ldots \frac{1}{x_l - i \Omega_n - i \omega_1 - \ldots - i \omega_{l-1}}
\end{gather*}
Considering summands of this expression as an analytic function of $\Omega_n$ we can decompose it as a sum of simple poles $\frac{C(\omega_i)}{x -i\Omega_n}$.
The summation over $\Omega_n$ yields
\begin{gather}
T \sum_n \frac{C(\omega_i)}{x - i \Omega_n} = \frac{C(\omega_i)}{2}\coth \frac{x}{2 T},\notag\\
\mathcal{A}_l\left(\omega_i\right) = \frac12 \int \prod^l_{i=1}\frac{dx_i}{2\pi} \Im G_R(x_i) \left[ \coth \frac{x_1}{2 T} \frac{1}{x_2 - x_1 - i \omega_1} \ldots \frac{1}{x_l - x_1 - i \omega_1 - \ldots - i \omega_{l-1}} +\right. \notag\\ \left. + \frac12 \coth \frac{x_2 - i \omega_1}{2 T}\frac{1}{x_1 - x_2 + i \omega_1} \ldots \frac{1}{x_l - x_2 - i \omega_2 - \ldots - i \omega_{l-1}}+\ldots\right]
\end{gather}
One can see that if $\sum^k_{i=l}\omega_i \neq 0$ for any choice of the indices the integral, the amplitude would not have divergences and the resulting function would be analytic.  Otherwise there are poles if some sum of the $\omega$'s go to zero. For instance, for $l=2$ we have
\begin{gather}
\mathcal{A}_2 (\omega_1) =  \frac12 \int \prod^l_{i=1}\frac{dx_i}{\pi} \Im G_R(x_i) \frac{\coth \frac{x_1}{2 T} + \coth \frac{x_2}{2 T}}{x_1 + x_2 - i \omega_1}.
\end{gather}
This formula is analogous to the susceptibility of the Fermi liquid \cite{abrikosov1962methods}, where such singularities lead to the different $k$ and $\omega$ limits. If $\omega_1 \neq 0$ the integral does not contain any singularities and therefore an analytic function of $\omega_1$. If $\omega_1=0$ there are possible divergences
\begin{gather}\label{therootofevil}
\lim_{x_1 \to -x_2}\frac{\coth \frac{x_1}{2 T} + \coth \frac{x_2}{2 T}}{x_1 + x_2 - i \omega_1} = \left\{\begin{matrix} 0,\quad \omega_1 \neq 0 \\ -\frac{1}{2 T \sinh^2 \frac{x_2}{2 T}},\quad \omega_1=0  \end{matrix}\right.
\end{gather}
This additional term can amount to the Kronecker anomaly at $\omega_1=0$. This singularity happens when $x_1 = -x_2$ and if the product of $\Im G_R(x_1)$ is not enough singular there would not be any contributions.  But if $\Im G_R(x_i)$ is singular, the term \eqref{therootofevil} will contribute and we would get a Kronecker anomaly.  
 In the example of the harmonic oscillator
\begin{gather}
\Im G_R(x_1) \Im G_R(x_2) = \pi^2 \delta(x_1^2 - m^2) \delta(x_2^2 - m^2),
\end{gather}
and two-dimensional integral reduces to a sum over the subset $x_1=\pm x_2=\pm m$. The additional term \eqref{therootofevil} amounts to the loop computation and coincides exactly with the discrepancy \eqref{paradox}. If we introduce a Kronecker function $\delta_{x,y}$ 
\begin{gather}
\Delta A_2(\omega_1)=  -\frac{1}{4 T} \int \prod^l_{i=1}\frac{dx_i}{\pi} \Im G_R(x_i) \frac{\delta_{x_1,x_2} \delta_{\omega_1,0}}{\sinh^2 \frac{x_2}{2 T}} \label{A:Step}
\end{gather}
We see only stable quasi-particles contributes to the Kronecker anomaly, when $\Im G_R$ has delta-functional singularities.
%
%
%
%
%

We present another way of understanding of Kronecker anomaly at $\omega=0$. Thus, we consider a two-point Whitman function of two operators $A(z=t+i\tau)$ and $B(0)$
\begin{gather}
    G_{AB}(z)=\frac{1}{Z} \tr\left[A(t+i\tau)B(0)e^{-\beta H}\right] = \label{eq:gab}\\
\frac{1}{Z} \sum_{n,m} \braket{n|A|m} \braket{m|B|n} e^{-\beta E_n} e^{(\tau+it) \left(E_n - E_m\right)},\notag
\end{gather}
Near the points $\tau=0,\beta$ this function could be rewritten as
\begin{gather}
    G_{AB}(t+i\beta)= \frac{1}{Z} \sum_{n,m} \braket{n|A|m} \braket{m|B|n} e^{-\beta E_m} e^{it \left(E_n - E_m\right)},\notag\\
    G_{AB}(t) = \frac{1}{Z} \sum_{n,m} \braket{n|A|m} \braket{m|B|n} e^{-\beta E_n}  e^{it \left(E_n - E_m\right)} = G^*_{AB}(t+i\beta)
\end{gather}
where one can notice that $G_{AB}(t+i\tau)$ is analytic function in the region $0<\tau<\beta$ (otherwise the sum over $n$ or $m$ would diverge). 
The analytical structure of this function is drawn in the fig. \eqref{fig:analgab}.   In the limit $z\to \infty$  the highly oscillating terms $e^{i t(E_n-E_m)}$ would cancel out all contributions unless $E_n=E_m$. Thus we define
\begin{gather}
    G_{AB}(\pm\infty) = g_{AB}= \frac{1}{Z} \sum_{n} \braket{n|A|n} \braket{n|B|n} e^{-\beta E_n} \in \mathbb{R}
\end{gather}
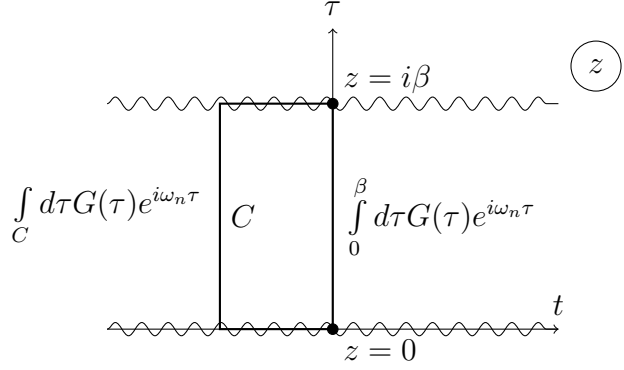
\begin{figure}
    \centering
    \begin{tikzpicture}
    \draw [decorate,decoration=snake] (-3,3) -- (3,3);
    \draw [decorate,decoration=snake] (-3,0) -- (3,0);
    \filldraw[black] (0,3) circle (2pt) node[anchor=south west] {$z=i\beta$};
    \filldraw[black] (0,0) circle (2pt) node[anchor=north west] {$z=0$};
    \draw[->] (-3,0)--(3,0)  node[ inner sep=5pt, pos=1, anchor=south]{$t$};
    \draw[->] (0,0)--(0,4) node[ inner sep=5pt, pos=1, anchor=south]{$\tau$};   
    \draw[thick] (0,0)--(0,3) node[ inner sep=5pt, pos=0.5, anchor=west]{$\int\limits^\beta_0 d\tau G(\tau)e^{i\omega_n \tau}$};
    \draw[thick] (0,0)--(-1.5,0)--(-1.5,3)--(0,3);
    \node at (-3,1.5) {$\int\limits_{C} d\tau G(\tau)e^{i\omega_n \tau}$};
    \node at (-1.5,1.5)[anchor=west]{$C$};
    \node at (3.5,3.5) [draw,circle]{$z$} ;
    \end{tikzpicture}
    \caption{Analytical structure of the two-point function \eqref{eq:gab}. The function $G_{AB}(z)$ has two cuts at $\Im z=0,\beta$.}
    \label{fig:analgab}
\end{figure}
The Fourier transform of $G_{AB}(\tau)$ is given by
\begin{gather}
    G_{AB}(\omega_n) = \int\limits^\beta_0 d\tau e^{i\omega_n \tau} G_{AB}(i\tau), 
\end{gather}
Let us consider the case of $\omega_n\leq 0$. Then we can deform the contour of integration \eqref{fig:analgab} and get
\begin{gather}
    G_{AB}(\omega_n) = \int\limits_C d\tau e^{i\omega_n \tau} G_{AB}(i\tau) = \notag\\
    =\lim_{t_0\to -\infty } e^{\omega_n t_0}\int\limits_C d\tau e^{i\omega_n \tau} G_{AB}(i\tau-t_0) + \int\limits^0_{-\infty} dt e^{\omega_n t} \Im G_{AB}(t) = I_n + \mathcal{F}(\omega_n),
\end{gather}
one can see that the second term $\mathcal{F}(\omega_n)$ is a analytical function of its argument $\omega_n$. The first term vanishes for $\omega_n < 0$ since it is multiplied by a bounded function. But taking into account that $G_{AB}\to g_{AB}$ at infinity we get
\begin{gather}
    I_n = \lim_{t_0\to -\infty } e^{\omega_n t_0}\int\limits_C d\tau e^{i\omega_n \tau} G_{AB}(i\tau-t_0) = \frac{g_{AB}}{T} \delta_{\omega_n,0},
\end{gather}
that is precisely a Kronecker anomaly.


\section{Two dimensional scalar electrodynamics on a torus} \label{sec:KAELGR}
In this section we show how Kronecker anomaly affects a dynamics of quantum electrodynamics and gravity coupled to a massive scalar and fermionic fields on torus $S^1_\beta \times S^1_{\beta'}$. It could be seen as a system at finite temperature with compact spatial coordinate. The action of this model is
\begin{gather}
S = \int d^2 x \left[ \left|\left(\partial_\mu - i e A_\mu\right) \phi \right|^2  - m^2 \left|\phi\right|^2 - \frac14 F_{01}^2\right].
\end{gather}
We can read out the propagator
\begin{gather}
G(n,m) = \frac{1}{p_n^2 +\omega_k^2+m^2},\quad \quad p_n = \frac{2\pi n}{\beta'},\quad \omega_k = \frac{2\pi k}{\beta}.
\end{gather}
The one-loop contribution to the polarization operator, i.e. in the lowest order of the coupling constant, $e^2$, 
\begin{gather*}
\Pi_{00}(\Omega,q) = \frac{e^2}{\beta'\beta}\sum_{n,m} \left[ \frac{\left(2 \omega_n +\Omega\right)^2}{(\omega_n^2 + p_m^2 + m^2) \left((\omega_n - \Omega)^2 + (p_m - q)^2 + m^2\right)} - \frac{2}{\omega_n^2 + p_m^2 + m^2}\right]
\end{gather*}
Using Ward identities we can find the following relations for the polarization operators
\begin{gather} \label{Wardident}
\Omega \Pi_{00}(\Omega,q) + q \Pi_{01}(\Omega,q) = 0,\quad \text{and}\quad  \Omega \Pi_{01}(\Omega,q) + q \Pi_{11}(\Omega,q) = 0,
\end{gather}
We note that if $\Omega \neq 0$ then we can express $\Pi_{00} = - \frac{q}{\Omega}\Pi_{01}$ and $\Pi_{00}(\Omega,q)$  must be equal to zero for $q=0$ and $\Omega \neq 0$. 
The equations \eqref{Wardident} do not constrain the polarization operators only when $\Omega=q=0$.

Therefore, non-analytical contributions can come only at $\Omega=q=0$. Thus we set $\Omega=0$ and consider the corresponding component as a function of the spatial momentum, $q$,
\begin{gather}
\Pi_{00}(0,q) = \frac{2 e^2}{\beta'\beta}\sum_{n,m} \left[ \frac{2 \omega_n^2}{ \left(\omega_n^2 + p_m^2 + m^2\right)  \left(\omega_n^2 + (p_m+q)^2 + m^2\right)} - \frac{1}{\omega_n^2 + p_m^2 + m^2}\right].
\end{gather}
Summing this series when $q=0$ and $q\neq 0$ we get the following result,
\begin{gather}
\Pi_{00}(0,q) =\pi_{00}\,\delta_{q,0} + \tilde{\Pi}_{00}(0,q), \quad \pi_{00} =  \frac{e^2}{\beta}\sum_n \frac{\omega_n^2}{\omega_n^2 + m^2} \frac{1}{ \sinh^2\left(\frac{\sqrt{\omega_n^2+m^2}}{2 T'}\right)}.
\end{gather}
The last term brings to the effective action of the electromagnetic field an additional non-trivial interaction for $A_0(0,0)$.

To understand this better we can study the $2d$ model with massless Dirac fermions on the same torus $S^1_\beta \times S^1_{\beta'}$ (with periodic boundary conditions over the spatial circle $S^1_{\beta'}$). In this case the propagator for the Kronecker mode is
\begin{gather}
\Pi_{00}(\omega=0,q) = e^2 T T' \sum_{n,m} \frac{2 \omega_n^2 - 2 p_m (p_m+q)}{\left(\omega_n^2 + p_m^2\right)\left(\omega_n^2 + (p_m+q)^2\right)}  = \notag\\ = \frac{1}{q} e^2 T' \sum_m \left(\tanh\left(\frac{p_m+q}{2T}\right) - \tanh\left(\frac{p_m}{2 T}\right)\right), \,\,
\text{here}\,\, T=\frac{1}{\beta}\,\, \text{and} \,\, T' = \frac{1}{\beta'}
\end{gather}
If $q\neq 0$ we can shift the summation and compute the difference between the first and second term in the summand to find that:
\begin{gather}
\Pi_{00}(\omega=0,q) =\frac{e^2}{2\pi}
\end{gather}
In the limit as $q\to 0$ we must replace the difference by the derivative to get that
\begin{gather}
\Pi_{00}(\omega=0,q=0)
 = e^2 \frac{\beta}{\beta'} \sum_n \frac{1}{\cosh^2\frac{\pi n \beta}{2\beta'}}  = e^2 g\left(\frac{\beta}{\beta'}\right)+\frac{e^2}{2\pi},\notag\\
\text{where}\quad g(x) = \sum^\infty_{n=-\infty} \frac{x}{\cosh^2 \frac{\pi n x}{2}}-\frac{1}{2\pi}.
\end{gather}
The sum is different again for $q=0$ and $q\neq 0$. 
We would like to stress that this additional term is not connected to the dynamics of the Wilson lines -- it is already captured by Schwinger term. Indeed, the Wilson line is equal to $W_t(x) = \int d\tau A_0(x,\tau)$ or in Fourier space $W_t(q) = A_0(0,q) = \frac{F_{01}(0,q)}{q}$ then we can rewrite the above contribution to the action as 
\begin{gather}
S = \frac{e^2}{2\pi}\sum_{q} \frac{F_{01}(0,q) F_{01}(0,-q)}{q^2} + e^2 g\left(\frac{\beta}{\beta'}\right)W^2_t(q=0) =\\= \sum_q W_t(q) W_t(-q)+ e^2 g\left(\frac{\beta}{\beta'}\right)W^2_t(q=0)  = \int dx\, W_t^2(x)+ e^2 g\left(\frac{\beta}{\beta'}\right) \left(\int dx\, dt\, A_0\right) ^2. \notag
\end{gather}
The action contains the local dynamics of the Wilson lines, while the last term does not have the same form and brings an additional dynamics. These non-local contributions at all orders of perturbation theory were firstly computed in \cite{Sachs:1991en}. 


An analogous calculations could be performed for the gravitational case. We will work with the linearized gravity in order to apply intuition we developed in the abelian case. Let us study a theory symmetric field $h_{ab}$ that is invariant under the following transformations
\begin{gather}\label{gravgauge}
h_{ab} \to h_{ab} + \partial_a \epsilon_b + \partial_b \epsilon_a, \quad h_{ab}(k) \to h_{ab}(k) + k_a \eps_b + k_b \eps_a,
\end{gather}
The Fourier transform of the eq. \eqref{gravgauge} degenerates for $\vec{k} = 0$, meaning that additional degrees of freedom could contribute to the dynamics. 
We couple this gravity theory to a fermion field on the torus $\mathbb{T}^2=S^1_\beta \times S^1_{\beta'}$. If we integrate out the fermions from the theory, these new degrees of freedom could become alive. 
The gravity field $h_{ab}$ is coupled to a stress energy tensor
\begin{gather}
\delta S = \int d^2 x h_{ab} T_{ab},\quad T_{00} = \bar{\psi} \gamma_0 \partial_0 \psi,\quad T_{01} = \frac12 \bar{\psi}\left[\gamma_0 \partial_1+\gamma_1 \partial_0\right]\psi, \quad T_{11} = \bar{\psi} \gamma_1 \partial_1 \psi.
\end{gather}
Due to the gauge invariance $\Pi_{00|00}(\Omega,q=0) = \braket{T_{00}(0,q) T_{00}(0,-q)} =0$  
when $\Omega \neq 0$.  Hence, to get the non-analytical contributions to the polarization operator $\Pi_{00|00}$ we set $\Omega=0$ and study it as a function of $q$.
The computation gives the following polarization operator 
\begin{gather}
\Pi_{00|00}(\Omega=0,q) = \frac{1}{\beta\beta'} \sum_{n,m}  \frac{\omega_n^2\left(\omega_n^2 - p_m(p_m+q)\right)}{(p_m^2+\omega^2_n)((p_m+q)^2+\omega^2_n)}.
\end{gather}
Summation over frequencies $\omega_n$ could be performed and it yields:
\begin{gather}
\Pi_{00|00}(\Omega=0,q) = \frac{1}{\beta'}\sum_n \frac{(p_n+q)^2 \tanh \frac{\beta(p_n+q)}{2}-p_n^2 \tanh \frac{\beta p_n}{2}}{2 q}
\end{gather}
This sum over $p_n$ is seemingly divergent but after regularization for $q \neq 0$:
\begin{gather}
\Pi_{00|00}(\Omega=0,q\neq 0) = -\frac{q^2}{6},
\end{gather}
This term brings to the effective action the following term
\begin{gather}
\delta S = h_{00}(\omega=0,q) \frac{q^2}{6} h_{00}(\omega=0,-q) = \frac{R(\omega=0,q) R(\omega,-q)}{6 q^2},
\end{gather}
where we have used that at linear order $R(\omega=0,q) = -q^2 h_{00}(\omega =0 ,q)$. Thus, we have managed to restore  Liouville action on a torus.  This fact is not surprising. 
But if we consider the case of $q=0$, when the constraints trivially satisfied, we get
\begin{gather}\label{gravAnom}
\delta\Pi_{00|00}(\Omega=0,q=0)=\sum_n \frac{p_n^2 + 2 \frac{p_n}{\beta}\sinh  \left[p_n \beta\right]}{\cosh^2\left[\frac{p_n \beta}{2}\right]}
\end{gather}
This term brings a non-local interaction to the effective action. 
It would be interesting to understand better the origin and the meaning of this term.

\section{Kronecker anomaly in spaces with constant curvature}\label{sec:KA3}
Before we considered only the cases of the spaces with compact subspace $\mathcal{M} = \tilde{\mathcal{M}} \times S^1$. The calculation was quite explicit and Kronecker anomalies appeared.  It would be interesting to study the case of other compact spaces. The most  direct generalization is to consider a three-dimensional sphere $S^3$. The problem is quite interesting and related to the problem of de Sitter stability \cite{Akhmedov:2013vka,Polyakov:2012uc}. Also it has already been shown that loop corrections  for principal series the loop corrections contain non-analytical contributions \cite{Akhmedov:2017ooy}. 
Let us study the following theory
\begin{gather}\label{dsfieldtheory}
S = \frac12 \int d^3 x \sqrt{-g} \left[\left(\partial_\mu a\right)^2 -m^2 a^2 + \lambda a^3\right],
\end{gather}
We pick the following metric on the sphere
\begin{gather}
ds^2 = d\phi^2 + \sin^2\phi\left(d\theta^2 + \sin^2 \theta \, d\varphi^2\right).
\end{gather}
Due to high symmetry of $S^3$ two point functions depend on the geodesic distance $Z=\cos\phi$ \cite{Polyakov:2012uc} 
\begin{gather}
	G_\mu (\cos \phi)  =  \frac{\sinh \mu \left(\pi  -\phi\right)}{\sinh \mu \pi \sin \phi}=\sum_{L=0}^\infty \frac{(L+1)}{L(L+2) + m^2} \frac{\sin (L+1) \phi}{\sin \phi},
\end{gather}
where $\mu = \sqrt{m^2-1}$ and $\frac{\sin (L+1)\phi}{\sin \phi}$ is a eigenfunction on the three-dimensional sphere. 
Let us note that 
\begin{gather}
	G_\mu(\cos \phi) = \sum_{K=-\infty}^\infty \frac{K}{K^2 + \mu^2} \frac{\sin K \phi}{\sin \phi},
\end{gather}
while to get Feynman propagator we should replace the summation with an integral
\begin{gather}
	Q_\mu(\cos \phi) = \int\limits^\infty_{-\infty} \frac{dK}{2\pi} \frac{K}{K^2 + \mu^2} \frac{\sin K \phi}{\sin \phi}
\end{gather}
We want to decompose the one-loop correction $\Pi(\cos \phi) = G_\mu^2(\cos\phi)$  through the eigenfunctions of the 3d sphere
\begin{gather} 
P_L(\cos \phi) =  \frac{\sin(L+1) \phi}{\sin \phi} = \sum_{K=L,L-2,\ldots,-L} e^{i K \phi}, \notag\\
\int\limits^\pi_0 d\phi \sin^2 \phi P_L(\cos\phi) P_K(\cos\phi) = \delta_{LK}, \notag\\ \Pi(\cos \phi) = \sum_L\Pi_L  P_L\left(\cos \phi\right), \quad
\Pi_L  = \int\limits^\pi_0 d\phi \sin^2 \phi \,\Pi(\cos \phi)\, P_L(\cos \phi),
\end{gather}
we get that
\begin{gather}
\Pi_{L} =\sum_{K=L,L-2,\ldots,-L}\pi_K, \quad \pi_K =  \frac{1}{\sinh^2 \pi \mu}\frac{ \mu \sinh 2\pi \mu}{K^2 + 4 \mu^2} - \frac{\pi}{2 \sinh^2 \pi \mu} \delta_{K,0},
\end{gather}
Summing this series over we get the following expression
\begin{gather}
    \Pi_L = - \frac{i}{4} \coth \pi \mu \left[H_{-i \mu - \frac{L}{2}-1}-H_{i \mu - \frac{L}{2}-1} - H_{-i \mu + \frac{L}{2}}+H_{i \mu + \frac{L}{2}}\right] -  \frac{\pi \delta_{L,{\rm even}}}{2\sinh^2 \pi \mu} 
\end{gather}
As one can see there is a non-analytical contribution at even $L$'s. If we sum them over, we would get that the Kronecker anomaly depends on $\phi$ 
\begin{gather}
  \delta\Pi(x,y)= \delta \Pi(\cos\phi)= -\frac{\pi}{2\sinh^2 \pi \mu} \sum_k \frac{\sin (2k+1)\phi}{\sin\phi} =  -\frac{\pi}{2\sinh^2 \pi \mu}  \frac{1}{\sin^2\phi} 
\end{gather}
Does it mean that the analytical continuation from $S^3$ to $dS^3$ is problematic? And if we change adiabatically  the mass, $m^2$, we would get isentropic change of the $\braket{a^2}$? 

As in the case of harmonic oscillator we can show that $\Pi_0$ corresponds to the change of the mass, either keeping temperature or entropy constant. Indeed,
\begin{gather}
G_\mu(\cos\phi) \approx \frac{1}{\phi} - \mu \coth \pi \mu + \mathcal{O}(\phi),\quad - \frac{\partial G_\mu(Z=1)}{\partial \mu^2} = \frac{\coth \pi \mu - \frac{\pi \mu}{\sinh^2 \pi \mu}}{2\mu} = \Pi_0
\end{gather}
On the other hand we can do this calculation in the $dS^3$ (that is done by analytical continuation $\phi \to i\alpha$) with the use of Keldysh-Schwinger technique
\begin{gather}
G_\mu(\cosh \alpha) =  i\frac{\sinh \mu \left(\pi  - i\alpha\right)}{\sinh \mu \pi \sinh \alpha}, \notag\\
- \frac{\partial G_\mu(Z=1)}{\partial \mu^2} = \int\limits^\infty_0 d\alpha \sinh^2 \alpha \operatorname{Im}\left[G^2_\mu(\cosh \alpha)\right]=  \frac {\coth \pi \mu}{2\mu},
\end{gather}
that is equal to the $\Pi_0$ without the Kronecker term. It means that if we change $\mu^2$ in the $dS_3$ instead of going to the BD vacuum we will descend in some other vacuum. Let us stress that we integrated over global $dS_3$ space. Therefore this result is not applicable for the Expanding Poincare Patch.  Nonetheless, we can study the Static Patch of de Sitter \cite{Akhmedov:2020qxd} which is related again to three-dimensional sphere by analytical continuation. It would be interesting to see whether such an non-analytical contribution to $\Pi_L$ would also relate the difference in the isothermal and isentropic susceptibilities. It shows, that the quantum field theory considered in the odd-dimensional de Sitter space does not make any sense --- the system is very sensitive to the way it was prepared. Any small adiabatic deformation of the theory could change its properties.

\subsection{Absence of Kronecker anomalies in even dimensional de Sitter spaces}
We want to generalize the previous calculation to  $S^2$. The propagator on a 2d sphere is
\begin{gather}
G(Z) = \frac{\pi}{\cosh\pi \mu} P_{-\frac12+i\mu}(-Z)
\end{gather}
Namely, we are interested in the decomposition of the self-energy function $\Pi(Z) = G(Z)^2$ as a sum over 2d sphere eigenfunctions
\begin{gather}
\Pi(Z) = \sum\limits^\infty_{L=0} \left(2L+1\right) P_L(Z) \Pi_L,
\end{gather}
where $P_L(Z)$ are Legendre polynomials. Using orthogonality we can extract coefficients 
\begin{gather}\label{eq:coef}
\Pi_L = \frac12 \int\limits^1_{-1} dZ \Pi(Z) P_L(Z),
\end{gather}this integral could be rewritten in the following way. First we notice that we have Legendre $Q$ function of integer index with the following property
\begin{gather}
\operatorname{Im} Q_n(x) = \frac{\pi}{2} P_n(x), \quad x \in\left[-1,1\right],
\end{gather}
then the eq. \eqref{eq:coef} is
\begin{gather}
\Pi_L = \frac{1}{\pi}\int\limits^1_{-1} dZ \Pi(Z) \operatorname{Im} Q_L(Z) =
 \frac{1}{2\pi i}\int\limits_{C} dZ \Pi(Z) Q_L(Z),
\end{gather}
where $C$ encircles the set $\left[-1,1\right]$, we deform this contour to $C_\infty=\lim\limits_{R\to \infty}\left\{Z \in \mathbb{C},|Z|=R\right\}$. Since $\Pi(Z) Q_L(Z) \sim \frac{1}{Z^{L+2}}$ for $L\geq 0$ 
\begin{gather}
\Pi_L =  \frac{1}{2\pi i}\int\limits_{C_\infty} dZ \Pi(Z) Q_L(Z) =   \frac{1}{\pi}\int\limits^\infty_1 dZ \operatorname{Im}\Pi(Z) Q_L(Z),
\end{gather}
but now $Q_L$ is a smooth function of $L$, namely $Q_L = Q_\nu e^{i\pi \nu}$, so we can introduce the following function
\begin{gather}
F(\nu) = \frac{1}{\pi}\int\limits^\infty_1 dZ \operatorname{Im}\Pi(Z) Q_\nu(Z) e^{i \pi \nu},
\end{gather}
this function is analytic and $F(L)=\Pi_L$ for integer $L$. We can check numerically that it is a nice agreement with the explicit resummation. Therefore, there are no Kronecker anomalies for even dimensional quantum field theories in de Sitter spaces.

\section{De Sitter Lacuna}
\label{sec:Lacuna}
In this section we will briefly discuss possible discontinuities of the propagator as a function of geodesic distance or interval in the background of non-trivial metric. The most simple example of such a background is again a de Sitter space. It is very well-known, that in de Sitter space due to external gravity, the in-vacuum is not stable. It could be seen by directly computing the decay rate
\begin{gather}
    \braket{\rm out| \rm in} = \int \mathcal{D}\phi e^{i \int d^d x \left[\frac{1}{2}(\partial_\mu\phi)^2 - \frac12 m^2 \phi^2 \right]}, \quad  \frac{\partial \log \braket{\rm out| \rm in}}{\partial m^2} = \int d^d x \braket{{\rm out}|\phi^2(x)| {\rm in}}
\end{gather}
The in-out matrix element could be computed in the following way. For instance in $d=2$, we would have the following in-out propagator
\begin{gather}
    \braket{{\rm out}|\phi(x)\phi(y)| {\rm in}} = Q_{-\frac12 + i \mu}(Z),\quad \mu = \sqrt{m^2 -\frac14},
\end{gather}
where $Z$ is a geodesic distance between these two points $x$ and $y$ in the de Sitter space. Then the imaginary part of the partition function is
\begin{gather}
    \operatorname{Im} \left[  \frac{\partial \log \braket{\rm out| \rm in}}{\partial \mu^2}\right] = \pi \tanh \pi \mu V_{dS},
\end{gather}
that suggests that in- and out- states in de Sitter space are orthogonal to each other. Therefore, we would expect if one turns on the interaction the in-state would decay to the out-state. It would be interesting to understand how the in-in propagator transforms to out-out propagator. In 2d, the in-in propagator could be obtained by a analytical continuation from sphere
\begin{gather}
    G_{\rm in/in}(Z) =\sum\limits^\infty_{L=0} \frac{(2L+1) P_L(Z)}{L(L+1)+m^2} = \frac{\pi}{\cosh \pi \mu} P_{-\frac12+i\mu}(-Z),
\end{gather}
the out-out propagator could not be computed by some analytical continuation.
Nevertheless, we can write down the expression for the out-out propagator in terms of modes
\begin{gather}
    G_{\rm out/out}(Z) = \int \frac{dp}{2\pi} e^{i p x} J_{i\mu}(p\eta) J_{-i\mu}(p\eta),
\end{gather}
we can see that the product of two Bessel functions, $J_{i\mu}(p\eta)J_{-i\mu}(p\eta)$ is an entire function of momentum $p$. Then if $|x| > 2\eta$ we can change the counter of integration to infinity
\begin{gather}
  \int \frac{dp}{2\pi} e^{i p x} J_{i\mu}(p\eta) J_{-i\mu}(p\eta) = 0, \quad |x|>2\eta \quad \Rightarrow \quad G_{\rm out/out}(Z) = \theta(Z+1) D(Z)
\end{gather}
Plugging this assumption into the equation for propagator we can see that $D(Z)$ should satisfy the same equation for massive scalar field in de Sitter space. And since $G_{\rm out/out}(Z)$ does not have singularities at $Z=-1$ we have
\begin{gather}
    G^{d=2}_{\rm out/out}(Z) = \theta(Z+1) G^{d=2}_{\rm in/in}(Z),
\end{gather}
we can find analogous formulae in higher dimensions
\begin{gather*}
    G^{d=2k+2}_{\rm out/out}(Z) = \frac{d^k}{dZ^k} G^{d=2}_{\rm out/out}(Z) = \theta(Z+1) G^{d=2k+2}_{\rm in/in}(Z) + C_1\delta(Z+1) + C_2 \delta'(Z+1) + \ldots 
\end{gather*}
In the case of the odd dimensions the situation a bit more strange. First, the in-out propagator does not contain imaginary contributions and therefore the in-state could not decay to the out-state \cite{Akhmedov:2019esv}. Second, the out-out propagator would be different
\begin{gather}
    G_{\rm out/out}(Z) = \int\frac{d^2 p}{(2\pi)^2} e^{i p x} J_{i\mu}(p\eta) J_{-i\mu}(p\eta) = \notag\\=\int \frac{p d p d\cos\theta_p}{2\pi} \frac{d\theta}{2\pi} \frac{d\theta'}{2\pi} e^{i\mu\left(\theta-\theta'\right)} e^{i p\left[x\cos\theta_p + \eta \cos(\theta-\theta_p)+\eta\cos(\theta'-\theta_p)\right]},
\end{gather}
from this we can see that if $|x| > 2\eta$ this integral is zero. For $|x| < 2\eta$ we can compute this integral
\begin{gather}
    G^{d=3}_{\rm out/out}(Z) = \theta(Z+1) \frac{e^{i\mu\theta}}{\sinh\theta} = \theta(Z+1) G_{\rm in/out}(Z),
\end{gather}
that is drastically different from the behavior in even dimensions. Again in higher dimensions
\begin{gather}
    G^{d=2k+3}_{\rm out/out}(Z) = \frac{d^k}{dZ^k}  G^{d=3}_{\rm out/out}(Z)= \theta(Z+1) G^{d=2k+3}_{\rm in/out}(Z)+C_1 \delta(Z+1) + \ldots
\end{gather}

\bibliographystyle{ssg}
\bibliography{Anomaly}

\begingroup\raggedright\begin{thebibliography}{10}

\bibitem{wilson1969non}
K.~G. Wilson, ``Non-Lagrangian models of current algebra,'' {\em Physical
  Review} {\bf 179} (1969), no.~5 1499.

\bibitem{kadanoff1969operator}
L.~P. Kadanoff, ``Operator algebra and the determination of critical indices,''
  {\em Physical Review Letters} {\bf 23} (1969), no.~25 1430.

\bibitem{polyakov1969properties}
A.~Polyakov, ``Properties of long and short range correlations in the critical
  region,'' {\em Zh. Eksp. Teor. Fiz} {\bf 57} (1969) 271--283.

\bibitem{larkin1969zh}
A.~Larkin and S.~Pikin, ``Zh. ETF 56 (1969) 1664; Soy. Phys,'' {\em JETP} {\bf
  29} (1969) 891.

\bibitem{pikin1993weak}
S.~Pikin, ``Weak first-order phase transitions,'' {\em Physica A: Statistical
  Mechanics and its Applications} {\bf 194} (1993), no.~1-4 352--363.

\bibitem{polyakov1991self}
A.~M. Polyakov, ``Self-tuning fields and resonant correlations in 2d-gravity,''
  {\em Modern Physics Letters A} {\bf 6} (1991), no.~07 635--644.

\bibitem{landau1986theory}
L.~D. Landau, E.~M. Lifshitz, A.~M. Kosevich, and L.~P. Pitaevskii, {\em Theory
  of elasticity: volume 7}, vol.~7.
\newblock Elsevier, 1986.

\bibitem{golubovic2000structural}
L.~Golubovi{\'c}, T.~Lubensky, and C.~O’hern, ``Structural properties of the
  sliding columnar phase in layered liquid crystalline systems,'' {\em Physical
  Review E} {\bf 62} (2000), no.~1 1069.

\bibitem{coleman1988there}
S.~Coleman, ``Why there is nothing rather than something: a theory of the
  cosmological constant,'' {\em Nuclear Physics B} {\bf 310} (1988), no.~3-4
  643--668.

\bibitem{hawking1990wormholes}
S.~W. Hawking, ``Do wormholes fix the constants of nature?,'' {\em Nuclear
  Physics B} {\bf 335} (1990), no.~1 155--165.

\bibitem{polchinski1998string}
J.~G. Polchinski, {\em String theory, volume I: An introduction to the bosonic
  string}.
\newblock Cambridge university press Cambridge, 1998.

\bibitem{Klebanov:1991hx}
I.~R. Klebanov and A.~M. Polyakov, ``{Interaction of discrete states in
  two-dimensional string theory},'' {\em Mod. Phys. Lett.} {\bf A6} (1991)
  3273--3281, \href{http://xxx.lanl.gov/abs/hep-th/9109032}{{\tt
  hep-th/9109032}}.

\bibitem{Polyakov:1991xa}
A.~M. Polyakov, ``{Singular states in 2-d quantum gravity},'' 9, 1991.

\bibitem{witten1992ground}
E.~Witten, ``Ground ring of two-dimensional string theory,'' {\em Nuclear
  physics B} {\bf 373} (1992), no.~1 187--213.

\bibitem{gross1991two}
D.~J. Gross, T.~Piran, and S.~Weinberg, {\em Two Dimensional Quantum Gravity
  And Random Surfaces-8th Jerusalem Winter School For Theoretical Physics},
  vol.~8.
\newblock World Scientific, 1991.

\bibitem{landau1980statisticheskaia}
L.~D. Landau and E.~M. Lifshits, {\em Statisticheskaia fizika}, vol.~5.
\newblock Pergamon, 1980.

\bibitem{abrikosov1962methods}
A.~Abrikosov, L.~Gor'kov, and I.~Dzyaloshinskii, ``Methods of Quantum Field
  Theory in Statistical Physics [in Russian],'' {\em GIFML, Moscow} (1962).

\bibitem{Sachs:1991en}
I.~Sachs and A.~Wipf, ``{Finite temperature Schwinger model},'' {\em Helv.
  Phys. Acta} {\bf 65} (1992) 652--678,
  \href{http://xxx.lanl.gov/abs/1005.1822}{{\tt 1005.1822}}.

\bibitem{Akhmedov:2013vka}
E.~Akhmedov, ``{Lecture notes on interacting quantum fields in de Sitter
  space},'' {\em Int. J. Mod. Phys. D} {\bf 23} (2014) 1430001,
  \href{http://xxx.lanl.gov/abs/1309.2557}{{\tt 1309.2557}}.

\bibitem{Polyakov:2012uc}
A.~Polyakov, ``{Infrared instability of the de Sitter space},''
  \href{http://xxx.lanl.gov/abs/1209.4135}{{\tt 1209.4135}}.

\bibitem{Akhmedov:2017ooy}
E.~Akhmedov, U.~Moschella, K.~Pavlenko, and F.~Popov, ``{Infrared dynamics of
  massive scalars from the complementary series in de Sitter space},'' {\em
  Phys. Rev. D} {\bf 96} (2017), no.~2 025002,
  \href{http://xxx.lanl.gov/abs/1701.07226}{{\tt 1701.07226}}.

\bibitem{Akhmedov:2020qxd}
E.~Akhmedov, K.~Bazarov, D.~Diakonov, and U.~Moschella, ``{Quantum fields in
  the static de Sitter universe},''
  \href{http://xxx.lanl.gov/abs/2005.13952}{{\tt 2005.13952}}.

\bibitem{Akhmedov:2019esv}
E.~T. Akhmedov, K.~V. Bazarov, D.~V. Diakonov, U.~Moschella, F.~K. Popov, and
  C.~Schubert, ``{Propagators and Gaussian effective actions in various patches
  of de Sitter space},'' {\em Phys. Rev.} {\bf D100} (2019), no.~10 105011,
  \href{http://xxx.lanl.gov/abs/1905.09344}{{\tt 1905.09344}}.

\end{thebibliography}\endgroup

\end{document}